\documentclass[reprint,amsmath,amssymb,aps,prl,superscriptaddress,nofootinbib]{revtex4-2}
\usepackage{graphicx}% Include figure files
\usepackage{dcolumn}% Align table columns on decimal point
\usepackage{bm}% bold math
\usepackage{color}
\usepackage{ulem}
\usepackage{comment}

\begin{document}

%\preprint{}

\title{Hierarchical structure of primary and hybridization-induced superconducting correlations in bilayer nickelates}


\author{Hiroshi Watanabe}
\email{watanabe.hiroshi@nihon-u.ac.jp}
\affiliation{
Department of Liberal Arts and Basic Sciences, College of Industrial Technology, Nihon University, Chiba 275-8576, Japan
}
\affiliation{
Research Organization of Science and Technology, Ritsumeikan University, Shiga 525-8577, Japan
}
\author{Hirofumi Sakakibara}
\affiliation{
Faculty of Engineering, Tottori University, Tottori 680-8552, Japan
}
\affiliation{
Advanced Mechanical and Electronic System Research Center (AMES), Tottori University, Tottori 680-8522,Japan
}
\author{Kazuhiko Kuroki}
\affiliation{
Department of Physics, The University of Osaka, Osaka 560-0043, Japan
}
%\thanks{}

%\collaboration can be followed by \email, \homepa-ge, \thanks as well.

\date{\today}

\begin{abstract}
High-pressure superconductivity in the bilayer nickelate La$_3$Ni$_2$O$_7$, with a transition temperature approaching 80 K, has stimulated intense debate regarding its microscopic origin.
Although an $s_{\pm}$ gap symmetry has been widely proposed, the electronic degrees of freedom responsible for pairing remain unsettled.
Here we investigate a bilayer two-orbital Hubbard model using the variational Monte Carlo method and reveal a hierarchical pairing structure in bilayer nickelates.
The primary pairing interaction originates from the bonding--antibonding splitting of the Ni $3d_{z^2}$ orbitals, while orbital hybridization redistributes superconducting correlations to the $d_{x^2-y^2}$ channel despite its weak intrinsic pairing interaction.
This distinction between the origin of pairing and resulting superconducting correlations explains why the two orbital channels exhibit comparable long-range correlations.
The resulting $s_{\pm}$ state is robust against changes in Fermi-surface topology.
These results reconcile apparently competing theoretical scenarios and provide a comprehensive understanding, highlighting the distinctive role of orbital hybridization in multilayer correlated superconductors.
\end{abstract}

%\keywords{}

\maketitle


% =========================
% Main text (no figures/captions)
% Paste into Overleaf as the manuscript body.
% =========================

%\section{Introduction}

Layered nickelates have recently emerged as a new platform for unconventional superconductivity, highlighted by the discovery of superconductivity in the bilayer compound La$_3$Ni$_2$O$_7$ under pressure with a transition temperature approaching 80~K~\cite{Sun2023sos}, comparable to that of cuprate superconductors.
Soon after the original discovery, zero resistance was attained in several studies~\cite{Zhang2024hsw,Wang2024psi}, confirming the occurrence of superconductivity.
This discovery has sparked intense interest in this field, and subsequent experiments revealed improved superconducting properties in related compounds, including enhanced superconducting volume fraction in La$_2$PrNi$_2$O$_7$~\cite{Wang2024bhs}, $T_c$ exceeding 90~K in (La,Sm)$_3$Ni$_2$O$_7$~\cite{Li2026bsu}, and ambient pressure superconductivity in thin films of La$_3$Ni$_2$O$_7$~\cite{Ko2025soa} and (La,Pr)$_3$Ni$_2$O$_7$~\cite{Zhou2025aso} with $T_c$ around 40~K.
%The latter has recently reached $T_c$ above 60~K~\cite{Zhou2025soa}.
%Superconductivity has also been observed in trilayer nickelates under pressure such as La$_4$Ni$_3$O$_{10}$~\cite{Sakakibara2024tao,Li2024sos,Zhu2024sip,Zhang2025sit} and Pr$_4$Ni$_3$O$_{10}$, with $T_c$ exceeding 20-30~K~\cite{Zhang2025bsi}.

Although La$_3$Ni$_2$O$_7$ shares a perovskite-derived layered structure with the cuprates, its electronic structure is markedly different, involving multiple Ni $3d$ orbitals and strong interlayer hybridization~\cite{Shilenko2023ces,Luo2023btm,Christiansson2023ces,Zhang2023esd,Cao2024fbp,Wu2024sac,Rhodes2024srt,Geisler2024sto,Wang2024eam,LaBollita2024asw} absent in single-layer cuprate systems.
These distinctions suggest that superconductivity in layered nickelates may arise from a microscopic mechanism fundamentally different from that of the cuprates, making La$_3$Ni$_2$O$_7$ an important test case for exploring high-temperature superconductivity beyond the cuprate paradigm.
In fact, even before the experimental discovery of superconductivity, one of the present authors considered La$_3$Ni$_2$O$_7$ as a possible candidate for realizing an extended $s$-wave superconducting state with a sign change between different electronic channels, commonly referred to as an $s_{\pm}$ state~\cite{Nakata2017fsf}.
Indeed, right after the experimental discovery, there appeared a number of theoretical studies proposing $s_{\pm}$- or interlayer $s$-wave pairing~\cite{Qin2023hsb,Yang2023pss,Huang2023iav,Shen2023ebm,Oh2023ttm,Lechermann2023eca,Liao2023eca,Qu2024btm,Kaneko2024pci,Sakakibara2024pht,Heier2024cda,Zhang2024spt,Lu2024ihs,Tian2024cea,Botzel2024tom,Kakoi2024pco,Chen2024osi,Luo2024hsi,Ryee2024qpb,Zhang2024spo,Yang2024spf,Schlomer2024sit,Pan2024eor,Xia2025sdo,Jiang2025top,Gu2025ema,Liu2025vqm,Qu2025hri,Ryee2025sgb,Gao2026rsp,Inoue2026umo,Le2025osa,Ushio2025tso,Maier2025ipi}.
However, despite this apparent agreement on symmetry, the microscopic origin of pairing and the factors controlling the superconducting transition temperature remain unresolved.

Prior theoretical studies of idealized bilayer Hubbard models have shown that bonding--antibonding splitting alone can stabilize an $s_{\pm}$ superconducting state with a high transition temperature~\cite{Kuroki2002hsi,Maier2011psa,Nomura2025shs}.
This has raised the intriguing possibility that La$_3$Ni$_2$O$_7$ may realize such a bilayer mechanism in a realistic material setting~\cite{Nakata2017fsf}.
At the same time, the actual electronic structure of La$_3$Ni$_2$O$_7$ involves substantial hybridization between the $3d_{z^2}$ and $3d_{x^2-y^2}$ orbitals (hereafter referred to as $z^2$ and $x^2-y^2$), rendering the situation more intricate than in minimal bilayer models.
It therefore remains unclear whether superconductivity in this compound can be understood within a simple bonding--antibonding framework or whether hybridization qualitatively reshapes the pairing mechanism.

A central source of this ambiguity lies in the origin of the $s_{\pm}$ state itself.
When only the Fermi surface is considered, $s_{\pm}$ pairing naturally emerges from spin fluctuations associated with nesting among the $\alpha$, $\beta$, and $\gamma$ Fermi-surface sheets~\cite{Yang2023pss,Lechermann2023eca,Heier2024cda,Zhang2024spt,Botzel2024tom,Luo2024hsi,Xia2025sdo,Jiang2025top,Gu2025ema}.
In this picture, $d$-wave or other pairing states compete with $s_{\pm}$, making the pairing symmetry sensitive to details of the Fermi-surface geometry.
Some studies further isolate the $\beta$ Fermi-surface sheet, whose topology resembles that of the high-$T_c$ cuprates~\cite{Jiang2024hsi,Fan2024sin}, and obtain a $d$-wave pairing state.
It should be noted, however, that the $\alpha$ and $\beta$ Fermi-surface sheets arise from hybridization between the $z^2$ and $x^2-y^2$ orbitals.
Without this hybridization the two bands would remain degenerate, so interpreting them as purely $x^2-y^2$-derived can be misleading.

On the other hand, a number of perturbative~\cite{Sakakibara2024pht,Le2025osa,Ushio2025tso,Ryee2025sgb,Gao2026rsp} and non-perturbative~\cite{Ryee2024qpb,Maier2025ipi,Liu2025vqm} approaches that incorporate the frequency dependence of the pairing interaction emphasize the importance of the bonding--antibonding $z^2$ bands split by interlayer coupling.
In these studies, the gap function or pair susceptibility exhibits a sign change between the bonding and antibonding $z^2$ bands, corresponding to interlayer intra-unit-cell pairing of the $z^2$ electrons with nearly momentum-independent gaps~\cite{Maier2025ipi,Ushio2025tso}.
This picture closely resembles the pairing mechanism proposed in bilayer Hubbard~\cite{Kuroki2002hsi} or $t$-$J$ models~\cite{Dagotto1992sil}, and several studies further report that the resulting $s_{\pm}$ pairing remains robust against changes in Fermi-surface topology such as the disappearance of the $\gamma$ sheet~\cite{Le2025osa,Ushio2025tso,Ryee2025sgb,Gao2026rsp}.

However, the situation is further complicated by studies suggesting that superconducting correlations are stronger in the $x^2-y^2$ orbital than in the $z^2$ orbital.
For example, a cellular dynamical mean-field theory (CDMFT) study reported a larger superconducting order parameter in the $x^2-y^2$ orbital~\cite{Tian2024cea}, whereas another CDMFT study obtained a smaller $x^2-y^2$ gap function than that of the $z^2$ orbital~\cite{Ryee2024qpb}.
Similar tendencies have also been obtained in several two-leg ladder studies that find stronger pair correlations in the $x^2-y^2$ orbital~\cite{Kaneko2024pci,Kakoi2024pco,Chen2024osi,Qu2025hri}, apparently contradicting the view that superconductivity is primarily driven by the $z^2$ orbital.

A related scenario treats the $z^2$ electrons as localized spins coupled by a strong interlayer exchange coupling $J_{\perp}$~\cite{Oh2023ttm,Qu2024btm,Zhang2024spo,Pan2024eor,Lu2024ihs,Yang2024spf,Schlomer2024sit}.
In this picture, an effective interlayer pairing interaction arises among the $x^2-y^2$ electrons when the Hund's coupling is sufficiently large, leading to $x^2-y^2$-derived superconductivity with interlayer $s$-wave pairing.
However, whether the $z^2$ electrons can indeed be regarded as localized spins should be carefully examined.
Indeed, it has been argued that such a Hund's-coupling-driven scenario may have difficulty accounting for experimentally observed non-Fermi-liquid behavior~\cite{Wang2025fla}.
It therefore remains unclear which electronic degrees of freedom drive superconductivity in La$_3$Ni$_2$O$_7$.

Adding to the above complexity, experimental studies remain divided on whether the purely $z^2$-derived $\gamma$ Fermi-surface sheet at the Brillouin-zone corner is present in thin films~\cite{Yang2024oec,Li2025aps,Sun2025oos,Wang2025eso}.
While the presence or absence of the $\gamma$ Fermi-surface sheet provides important information about the electronic structure, its absence does not necessarily imply that the $z^2$ orbital plays no role in superconductivity.
This is because the $\alpha$ and $\beta$ Fermi-surface sheets also contain substantial $z^2$ character due to strong orbital hybridization.

Thus, theories that nominally predict the same gap symmetry can rely on qualitatively distinct electronic degrees of freedom, obscuring which bands are essential for superconductivity and what ultimately determines the transition temperature. 

In this work, we investigate superconductivity in La$_3$Ni$_2$O$_7$ using variational Monte Carlo calculations~\cite{Ceperley1977mcs,Yokoyama1987vms} applied to a bilayer two-orbital Hubbard model .
This non-perturbative approach allows us to analyze both the superconducting gap structure and long-range superconducting correlations within a unified framework.

We demonstrate that superconductivity in La$_3$Ni$_2$O$_7$ exhibits a hierarchical structure of pairing.
The primary pairing interaction originates from the bonding–antibonding splitting of the $z^2$ bands, while orbital hybridization redistributes superconducting correlations to the $x^2-y^2$ channel without generating a sizable intrinsic pairing interaction in that orbital.
The resulting gap structure exhibits a clear $s_{\pm}$ character, with comparable superconducting correlations in the $z^2$ and $x^2-y^2$ channels.
These features are insensitive to the Fermi-surface topology, including the presence or absence of the $\gamma$ Fermi-surface sheet.
This hierarchical structure resolves the apparent tension between competing theoretical scenarios and provides a comprehensive understanding on how superconductivity in La$_3$Ni$_2$O$_7$ emerges from electronic degrees of freedom beyond those directly visible at the Fermi level.
Our results highlight the importance of bilayer structure and orbital hybridization in stabilizing superconductivity and point to the broader relevance of interlayer orbital physics in correlated superconductors.

\section{Results}
\subsection{Model and band structure}

\begin{figure}
\centering
\includegraphics[width=0.8\hsize]{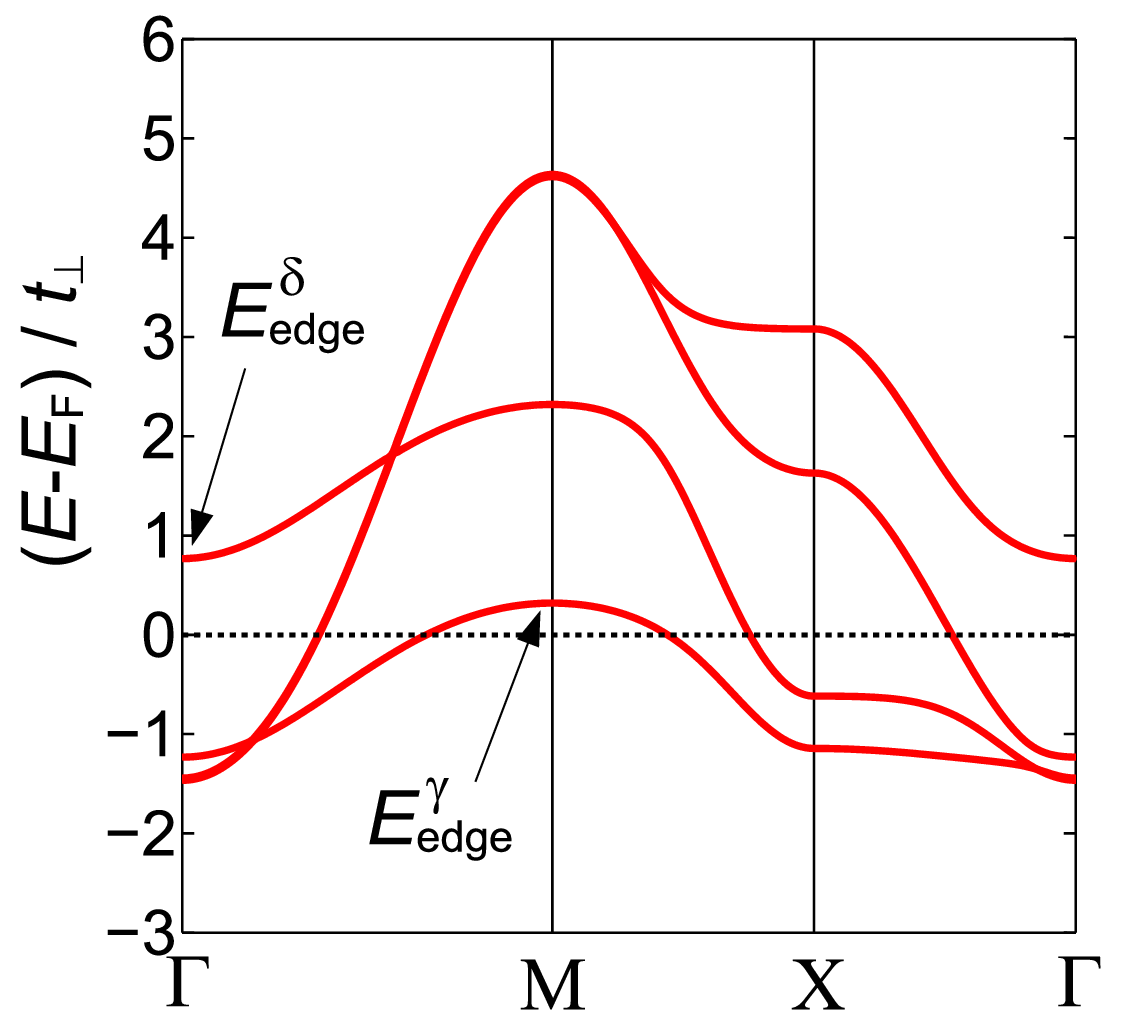}
\caption{
\label{band}
Noninteracting band structure of the bilayer two-orbital Hubbard model for $\Delta E/t_{\perp}=0.50$.
High-symmetry points are labeled as $\Gamma(0,0)$, M$(\pi,\pi)$, and X$(\pi,0)$.
The four bands originate from the $z^2$ and $x^2-y^2$ orbitals with interlayer hybridization.
The Fermi level is indicated by the horizontal dashed line.
The bonding and antibonding $z^2$ bands are denoted as $\gamma$ and $\delta$, respectively.
The quantities $E^{\gamma}_{\mathrm{edge}}$ and $E^{\delta}_{\mathrm{edge}}$ denote the band edges of the $\gamma$ and $\delta$ bands measured from the Fermi level, defined as the extrema closest to $E_F$.
The energy difference $E^{\delta}_{\mathrm{edge}}-E^{\gamma}_{\mathrm{edge}}$ remains unchanged as $\Delta E/t_{\perp}$ is varied.}
\end{figure}

To capture the essential electronic structure of La$_3$Ni$_2$O$_7$, we consider a bilayer two-orbital Hubbard model comprising the $z^2$ and $x^2-y^2$ orbitals.
The noninteracting part of the Hamiltonian is constructed to reproduce the band structure obtained from first-principles calculations for La$_3$Ni$_2$O$_7$ under the pressure of 20~GPa~\cite{Ochi2025tso}. 
Figure~\ref{band} shows the energy dispersion of the present four band model.
The key parameter controlling the relative orbital hierarchy is the ratio $\Delta E/t_\perp$, where $\Delta E=\varepsilon_{x^2-y^2}-\varepsilon_{z^2}$ denotes the level offset between orbitals, and $t_\perp$ the interlayer hopping amplitude of $z^2$ electrons.

For parameters directly extracted from density functional theory, $\Delta E/t_\perp \approx 0.50$, under which superconductivity is not stabilized within the present variational framework.
However, a previous study that goes beyond LDA calculations effectively enhances the separation between the orbital levels~\cite{Wang2024eam}, corresponding to a larger effective $\Delta E/t_\perp$.
Motivated by this observation, we treat $\Delta E/t_\perp$ as a physically relevant tuning parameter and systematically explore its impact on superconductivity while keeping all other hopping parameters fixed.
This procedure preserves the bonding--antibonding splitting structure; in particular, the energy difference $E^{\delta}_{\mathrm{edge}}-E^{\gamma}_{\mathrm{edge}}$ remains unchanged (see Fig.~\ref{band}).
It therefore allows us to isolate the role of the orbital hierarchy in driving superconductivity.
In this sense, varying $\Delta E/t_\perp$ effectively mimics a correlation-induced renormalization of the orbital hierarchy.
Electron--electron interactions are incorporated via intraorbital and interorbital Coulomb interactions, Hund's coupling, and pair-hopping terms, and the ground state is obtained using variational Monte Carlo calculations.
Details of the Hamiltonian, variational wave function, and numerical procedures are provided in Methods.

Unless otherwise specified, the Coulomb interaction parameters are fixed to $(U, U', J, J') = (8.0, 5.6, 1.2, 1.2)\, t_\perp$ with $t_\perp=0.659$~eV, satisfying the rotational symmetry condition $U' = U - 2J$ and $J = J'$.
The calculations are performed on a $2\times2\times L\times L$ lattice with $L=20$.
The total electron density is fixed at $n_{z^2}+n_{x^2-y^2}=1.5$, corresponding to a total of $N_{\text{e}}=1200$ electrons for $L=20$.
We have also performed calculations for $L=16$ and $L=24$ (for selected parameters) and confirmed that the qualitative physics remains unchanged.

\subsection{Interaction-driven reorganization of the electronic structure}

\begin{figure}[t]
\centering
\includegraphics[width=1.0\hsize]{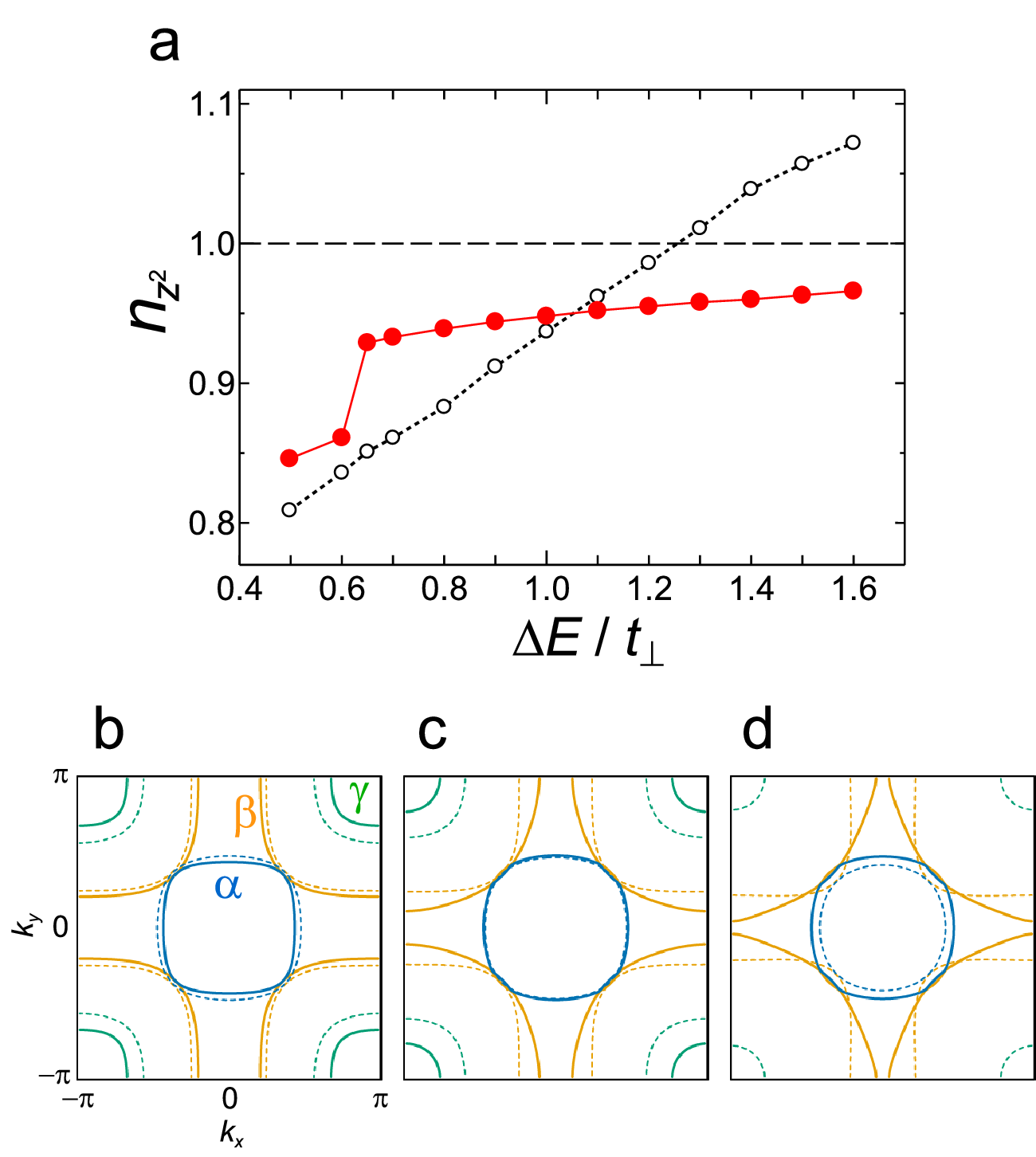}
\caption{
\label{nd2}
Interaction-driven reorganization of orbital occupancy and Fermi-surface topology.
(a) $z^2$ orbital occupancy $n_{z^2}$ as a function of $\Delta E/t_\perp$.
Open symbols (dashed line) denote the noninteracting case, while filled symbols (solid line) include the effect of Coulomb interactions.
In the interacting case, $n_{z^2}$ is strongly renormalized and becomes nearly pinned over a broad parameter range, indicating an interaction-induced redistribution of charge between orbitals.
The horizontal dashed line marks half filling of the $z^2$ orbital.
(b--d) Evolution of the Fermi-surface sheets for representative values of $\Delta E/t_\perp$:
(b) $\Delta E/t_\perp=0.50$ (non-superconducting regime),
(c) $\Delta E/t_\perp=0.65$,
and (d) $\Delta E/t_\perp=1.20$.
Solid lines correspond to the interacting case and dashed lines to the noninteracting case.
The $\alpha$, $\beta$, and $\gamma$ sheets are defined in (b).
With increasing $\Delta E/t_\perp$, the $\gamma$ sheet shrinks substantially and eventually disappears, while the $\beta$ sheet is strongly reshaped.
}
\end{figure}

We first examine how the electronic structure evolves as a function of the ratio $\Delta E/t_\perp$, which controls the relative energy offset between the two orbitals and the interlayer hopping amplitude of $z^2$ electrons.
As shown in Fig.~\ref{nd2}(a), in the noninteracting case, the $z^2$ orbital occupancy $n_{z^2}$ increases almost linearly with increasing $\Delta E/t_\perp$, indicating that $\Delta E/t_\perp$ provides a well-defined control parameter.
In contrast, once interactions are included, $n_{z^2}$ is strongly renormalized and becomes nearly pinned over a broad parameter range for $\Delta E/t_\perp \gtrsim 0.65$, where superconductivity is stabilized.

Because the total electron density is fixed at $n_{z^2}+n_{x^2-y^2}=1.5$, this behavior reflects an interaction-driven redistribution of charge between orbitals, resulting in a reorganization of the low-energy electronic structure.
Notably, $n_{z^2}$ is renormalized to values close to, but slightly below, half filling.
In this regime, electronic correlations remain substantial while finite carrier mobility is retained, providing a favorable balance between correlation strength and itinerancy for superconductivity.

This reorganization is also visible in the Fermi-surface sheets.
We label the three sheets as $\alpha$, $\beta$, and $\gamma$, as shown in Fig.~\ref{nd2}(b).
In the non-superconducting regime, the renormalization of the Fermi surface is relatively weak, and the $\gamma$ sheet becomes slightly smaller [Fig.~\ref{nd2}(b), $\Delta E/t_\perp = 0.50$].
In contrast, in the superconducting regime, the $\gamma$ sheet shrinks substantially and eventually disappears for $\Delta E/t_\perp \gtrsim 1.20$, as shown in Figs.~\ref{nd2}(c) and (d).
The $\beta$ sheet is also strongly renormalized and changes its shape noticeably.
Such drastic changes in the Fermi surface are also observed in a recent CDMFT study~\cite{Ryee2024qpb}, indicating the importance of strong correlation effects.
In the following, we examine how the superconducting pairing strength and correlation functions respond to these modifications of the Fermi-surface topology.

\subsection{Variational superconducting gap in the bonding--antibonding $z^2$ channel}

\begin{figure}[t]
\centering
\includegraphics[width=1.0\hsize]{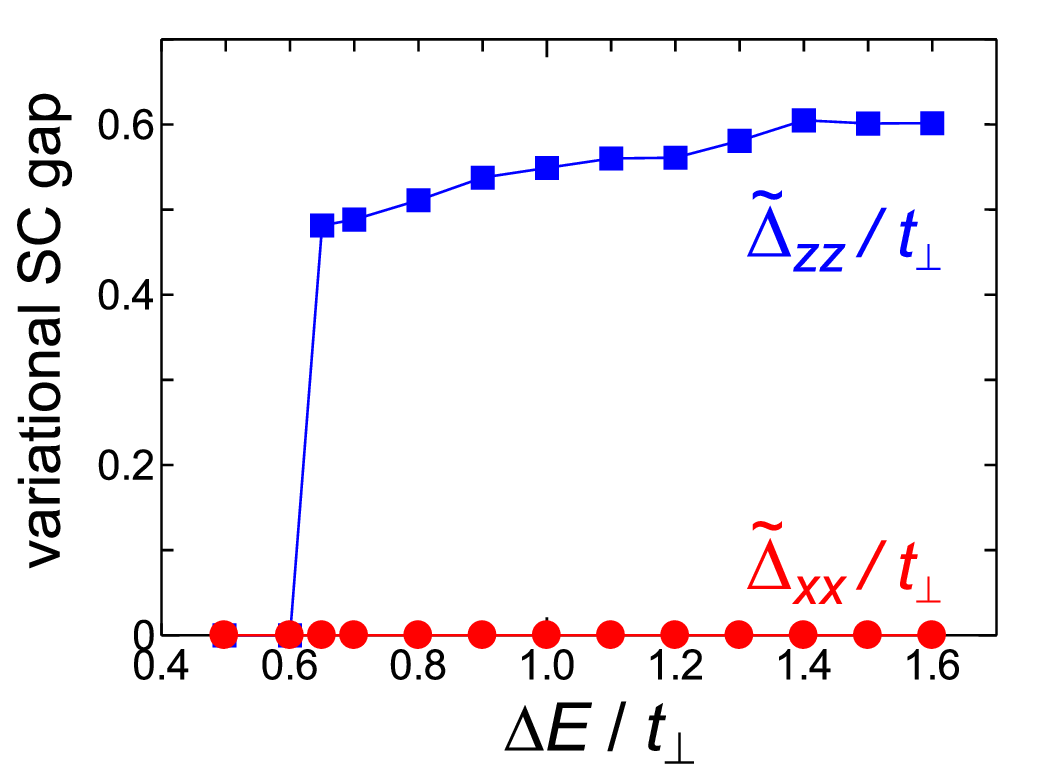}
\caption{
\label{Delta}
Variational superconducting gap parameters as a function of $\Delta E/t_\perp$.
The interlayer $z^2$ gap $\tilde{\Delta}_{zz}$ becomes finite for $\Delta E/t_\perp \gtrsim 0.65$ and increases gradually with increasing $\Delta E/t_\perp$, indicating that the primary pairing interaction develops in the bonding--antibonding $z^2$ channel.
In contrast, the interlayer $x^2-y^2$ gap $\tilde{\Delta}_{xx}$ remains negligible over the entire parameter range, demonstrating that pairing is not generated primarily in the $x^2-y^2$ orbital.
}
\end{figure}

We next analyze the variational superconducting gap parameters shown in Fig.~\ref{Delta}.
To explicitly indicate that these are variational parameters, we denote them by $\tilde{\Delta}_{zz}$ and $\tilde{\Delta}_{xx}$.
The interlayer $z^2$ gap $\tilde{\Delta}_{zz}$ becomes finite for $\Delta E/t_\perp \gtrsim 0.65$, concomitantly with the enhancement of $n_{z^2}$, indicating that the primary pairing interaction develops in the bonding--antibonding $z^2$ channel.
In contrast, although pairing in the $x^2-y^2$ channel is allowed in the variational space, the corresponding variational parameter $\tilde{\Delta}_{xx}$ consistently relaxes to negligible values upon optimization and is therefore effectively zero across the parameter range.

We emphasize that $\tilde{\Delta}_{xx} \approx 0$ as a variational parameter does not necessarily imply the absence of superconducting correlations in the $x^2-y^2$ channel.
In a BCS-type variational framework, the gap parameter can be viewed schematically as $\tilde{\Delta} \sim V_{\mathrm{eff}}\langle c c \rangle$, where $V_{\mathrm{eff}}$ is the effective pairing interaction and $c$ the electron annihilation operator.
Thus, $\tilde{\Delta}_{xx} \approx 0$ signifies the absence of a relevant attractive interaction in the $x^2-y^2$ channel rather than the absence of superconducting correlations.
This is consistent with the electronic structure: the $x^2-y^2$ orbital is primarily extended within each layer and has negligible interlayer hopping, whereas the $z^2$ orbital exhibits strong bonding--antibonding splitting that naturally promotes pairing.

\subsection{Long-range superconducting correlations in both $z^2$ and $x^2-y^2$ channels}
\begin{figure}
\centering
\includegraphics[width=1.0\hsize]{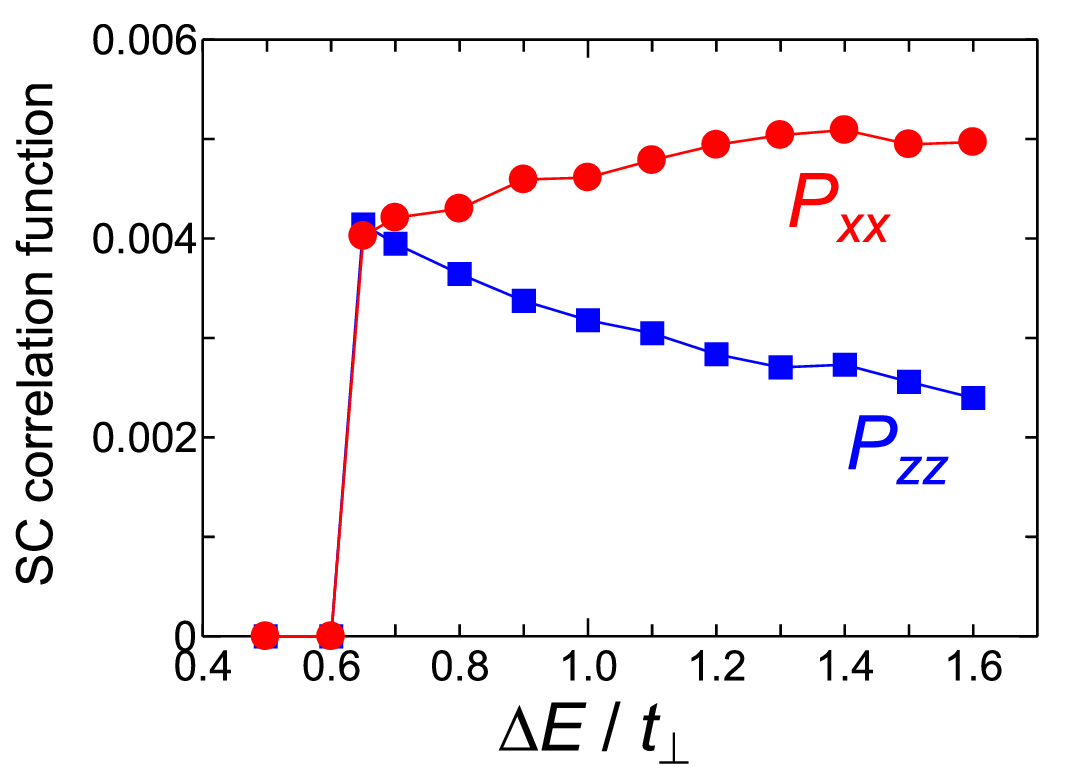}
\caption{
\label{PSC}
Long-range superconducting correlation functions as a function of $\Delta E/t_\perp$.
$P_{zz}$ and $P_{xx}$ denote the pairing correlations in the $z^2$ and $x^2-y^2$ channels, respectively.
$P_{xx}$ exceeds $P_{zz}$ over a broad parameter range and exhibits a contrasting dependence on $\Delta E/t_\perp$.
This behavior indicates that, although the effective pairing interaction is strongest in the $z^2$ channel, orbital hybridization redistributes superconducting correlations to the $x^2-y^2$ channel.
}
\end{figure}

Despite the dominance of $\tilde{\Delta}_{zz}$, the long-range superconducting correlations exhibit a markedly different evolution.
Figure~\ref{PSC} shows the $\Delta E/t_\perp$ dependence of the superconducting correlation functions $P_{zz}$ and $P_{xx}$ (see Methods for their definitions), which characterize the strength of long-range superconducting correlations in the $z^2$ and $x^2-y^2$ orbital channels, respectively.
Both $P_{zz}$ and $P_{xx}$ become finite for $\Delta E/t_\perp \gtrsim 0.65$, signaling the onset of superconductivity.
Remarkably, $P_{xx}$ becomes comparable to, and in part slightly larger than, $P_{zz}$ despite the absence of a sizable optimized $\tilde{\Delta}_{xx}$.
While the variational parameter $\tilde{\Delta}$ reflects the strength of a local pairing field within the trial wave function, $P$ captures the spatial propagation of superconducting correlations and is therefore indicative of superconducting order.
In the present case, although the effective pairing interaction is strongest in the $z^2$ channel, orbital hybridization redistributes superconducting correlations across orbital channels, resulting in comparable long-range correlations in both.
These results are reminiscent of a DMRG study on a two-orbital, two-leg $t$-$J$ ladder by two of the present authors and coworkers, where the $x^2-y^2$ interchain pair correlation is found to be quasi-long ranged through interorbital hybridization, even when Hund's coupling and $x^2-y^2$ interchain exchange coupling are turned off~\cite{Kakoi2024pco}.

Interestingly, a recent CDMFT study examining the filling dependence of orbital-resolved pairing amplitudes~\cite{Tian2024cea} reported a similar trend, where the pairing amplitude in the $x^2-y^2$ channel exceeds that in the $z^2$ channel and exhibits an opposite evolution toward half filling.
Although the calculated quantities and control parameters differ from those in the present work, the qualitative consistency supports the robustness of the underlying orbital differentiation.

\subsection{Orbital-resolved density of states}
\begin{figure}
\centering
\includegraphics[width=1.0\hsize]{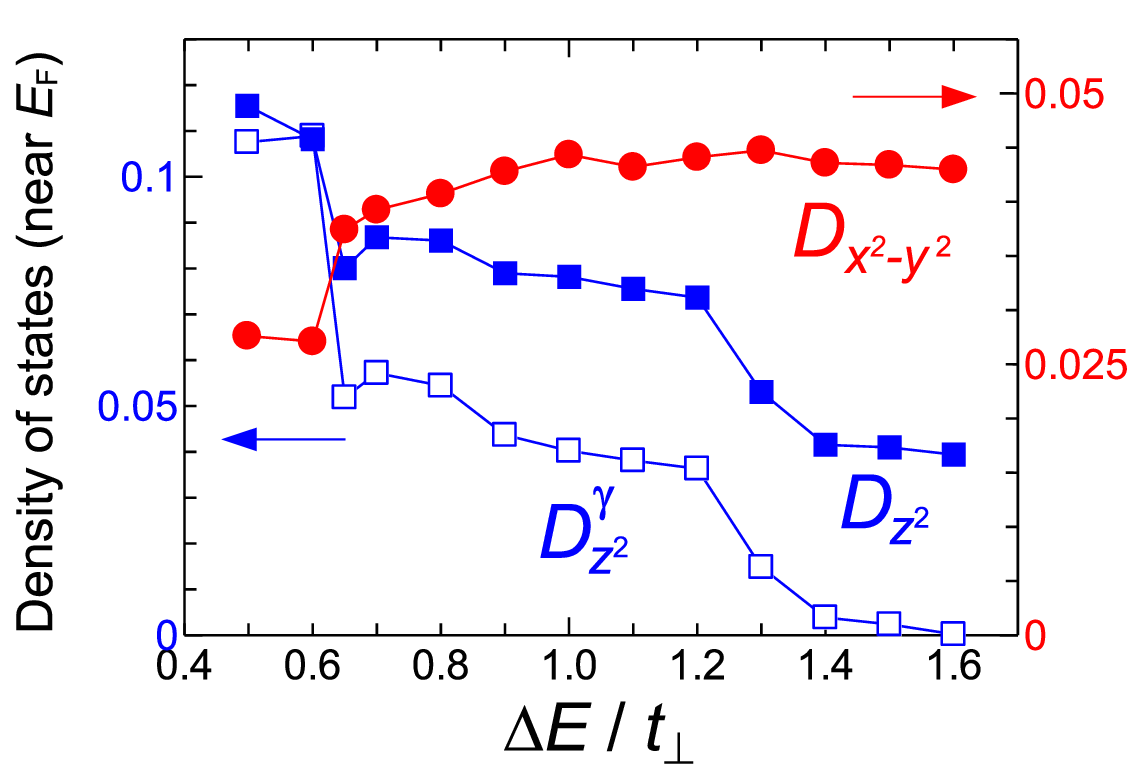}
\caption{
\label{DOS}
Orbital- and band-resolved density of states near the Fermi level as a function of $\Delta E/t_\perp$.
Open blue squares represent the $z^2$ density of states associated with the $\gamma$ band, $D^{\gamma}_{z^2}$, while filled blue squares denote the total $z^2$ density of states $D_{z^2}$ obtained by summing over all bands.
Red circles show the $x^2-y^2$ density of states $D_{x^2-y^2}$, which mainly originates from the $\alpha$ and $\beta$ bands.
As $\Delta E/t_\perp$ increases, $D^{\gamma}_{z^2}$ decreases as the $\gamma$ Fermi-surface sheet shrinks and disappears, whereas $D_{z^2}$ remains finite due to the hybridization-induced $z^2$ component on the $\alpha$ and $\beta$ bands.
The evolution of these quantities correlates with the behavior of the superconducting correlation functions shown in Fig.~\ref{PSC}.
}
\end{figure}

To clarify the origin of the behavior of $P_{zz}$ and $P_{xx}$, we examine the orbital- and band-resolved density of states near the Fermi level $D^{\mu}_{m}$ (see Methods for their definitions).
The orbital-resolved density of states $D_{m}$, obtained by summing over bands, is also considered.
Figure~\ref{DOS} shows the evolution of these quantities as a function of $\Delta E/t_\perp$.

As $\Delta E/t_\perp$ increases, the $z^2$ density of states associated with the $\gamma$ band, $D^{\gamma}_{z^2}$ (blue open squares), decreases as the $\gamma$ Fermi-surface sheet shrinks and eventually disappears.
However, because of strong hybridization between the $z^2$ and $x^2-y^2$ orbitals, a substantial $z^2$ component remains on the $\alpha$ and $\beta$ bands.
As a result, the total $z^2$ density of states, $D_{z^2}$ (blue filled squares), decreases but does not vanish even after the $\gamma$ Fermi-surface sheet disappears.

In contrast, the $x^2-y^2$ density of states $D_{x^2-y^2}$, which is carried almost entirely by the $\alpha$ and $\beta$ bands, increases with $\Delta E/t_\perp$ and eventually saturates.
The evolution of these orbital-resolved densities of states closely correlates with the behavior of the superconducting correlations shown in Fig.~\ref{PSC}:
while $P_{xx}$ grows together with the increasing $x^2-y^2$ density of states, $P_{zz}$ decreases but remains finite because the $z^2$ spectral weight near the Fermi level persists on the $\alpha$ and $\beta$ bands.

These results indicate that the magnitude of superconducting correlations is governed by the orbital character of the low-energy density of states, even though the pairing interaction itself originates from the bonding--antibonding $z^2$ channel.

\subsection{Gap structure in momentum space}

\begin{figure}
\centering
\includegraphics[width=0.9\hsize]{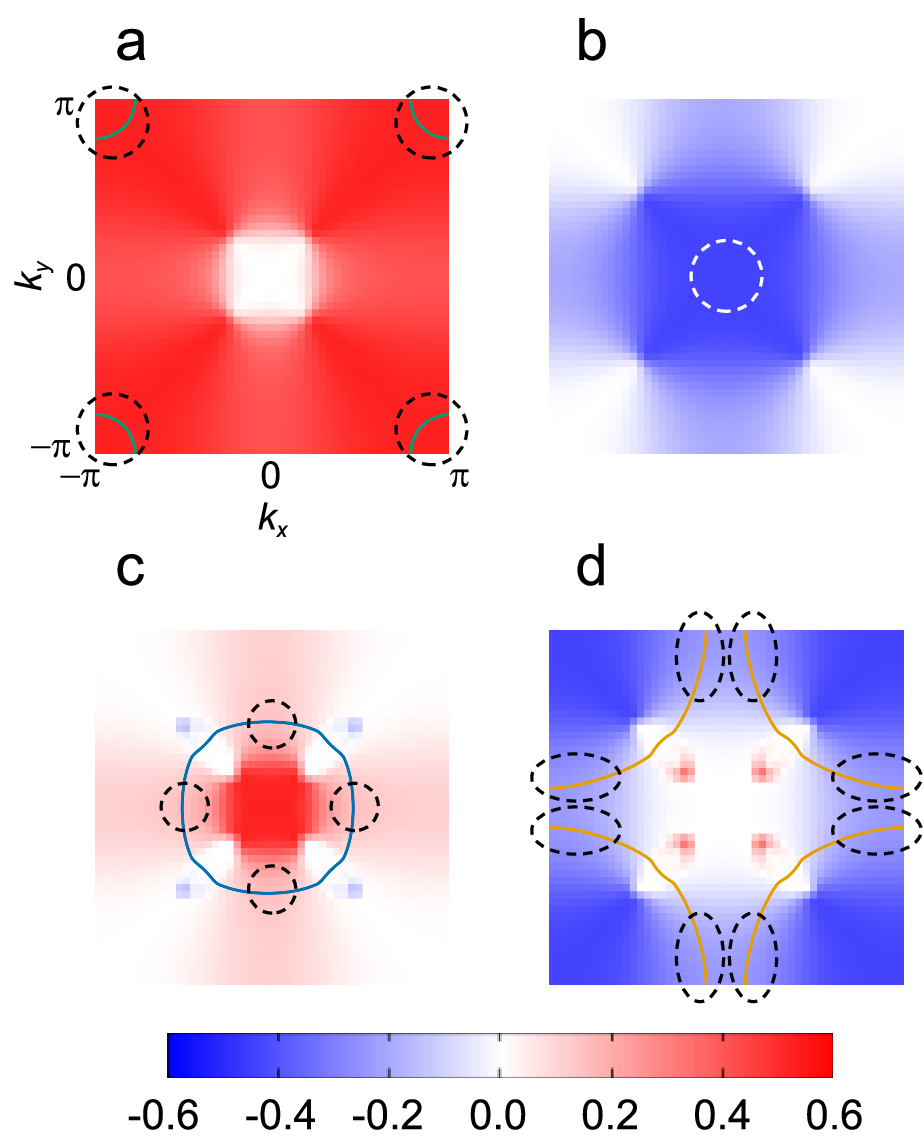}
\caption{
\label{gap}
Superconducting gap structure in the band representation for $\Delta E/t_\perp = 0.65$.
(a–d) Gap amplitude on the $\gamma$, $\delta$, $\alpha$, and $\beta$ bands, respectively.
The solid curves in (a), (c), and (d) represent the corresponding Fermi-surface sheets.
The $\gamma$–$\delta$ channel exhibits large and nearly momentum-independent gaps characteristic of the bonding–antibonding $z^2$ pairing channel.
In contrast, the gaps on the $\alpha$ and $\beta$ bands show strong momentum dependence and vanish along the $k_x=\pm k_y$ lines where the hybridization is forbidden by symmetry.
Dashed circles indicate representative regions discussed in the text.
Note that the bands are defined according to the energy ordering; apparent discontinuities therefore arise at band crossings where the band indices interchange.
}
\end{figure}

To visualize the momentum-space structure of the superconducting state, we plot the superconducting gap in the band representation in Fig.~\ref{gap} for $\Delta E/t_\perp=0.65$.
The resulting gap structure exhibits a clear $s_{\pm}$ character, with opposite signs between the $\gamma$ and $\delta$ bands and between the $\alpha$ and $\beta$ bands.

A notable feature is that the largest and nearly uniform gap amplitude appears in the $\gamma$-$\delta$ channel.
On the $\gamma$ band, the gap becomes large around M$(\pi,\pi)$ [dashed circles in Fig.~\ref{gap}(a)], where the $z^2$-derived $\gamma$ Fermi-surface sheet is located.
On the $\delta$ band, a large gap appears around $\Gamma(0,0)$ [dashed circle in Fig.~\ref{gap}(b)], where the Fermi surface is absent but the $z^2$-derived incipient band lies slightly above the Fermi level.
Overall, the gaps on the $\gamma$ and $\delta$ bands are nearly momentum independent and substantially larger than those on the $\alpha$ and $\beta$ bands.
This indicates that the primary pairing interaction originates in the bonding–antibonding $z^2$ channel.

In contrast, the gaps on the $\alpha$ and $\beta$ bands exhibit strong momentum dependence.
They become large in regions where the $z^2$ orbital strongly hybridizes with the $x^2-y^2$ orbital [dashed circles in Figs.~\ref{gap}(c) and (d)], while they vanish along the $k_x=\pm k_y$ lines where the hybridization is forbidden by symmetry.
This behavior demonstrates that the superconducting gap on the $\alpha$ and $\beta$ bands is induced through orbital hybridization rather than originating from an intrinsic pairing interaction in the $x^2-y^2$ channel.

Importantly, this gap structure remains essentially unchanged even when the $\gamma$ Fermi-surface sheet disappears at larger $\Delta E/t_\perp$.
The pairing state is therefore not sensitive to the detailed topology of the Fermi surface.
Instead, it is governed by the orbital composition of the low-energy electronic states, which remains largely unchanged on the $\alpha$ and $\beta$ bands due to persistent $z^2$ hybridization.

The obtained gap structure and the robustness of the superconductivity are consistent with previous studies that incorporates the frequency dependence of the pairing interaction~\cite{Le2025osa,Ushio2025tso,Ryee2025sgb,Gao2026rsp}, indicating the importance of finite-energy contribution.

\section{Discussion}
Our results establish a comprehensive understanding of superconductivity in La$_3$Ni$_2$O$_7$ rooted in the bonding--antibonding $z^2$ channel of its bilayer electronic structure.
Although various theoretical studies have proposed an $s_{\pm}$ gap symmetry, the microscopic origin of pairing has remained unsettled.
We demonstrate that the primary pairing interaction arises from the bonding--antibonding splitting of the $z^2$ orbital, while superconducting correlations in the $x^2-y^2$ orbital are induced indirectly through orbital hybridization.

A central outcome of this work is the separation between the origin of pairing and the resulting long-range superconducting correlations.
In the band representation, the optimized state exhibits a clear $s_{\pm}$ structure, with sign changes both between the $\gamma$ and $\delta$ bands and between the $\alpha$ and $\beta$ bands.
Importantly, the largest and nearly momentum-independent gap appears in the $\gamma$-$\delta$ channel, indicating that the bonding--antibonding $z^2$ splitting provides the primary driving force of pairing, whereas the $\alpha$–$\beta$ sign reversal emerges as a consequence of orbital hybridization.
The strong momentum dependence of the $\alpha$ and $\beta$ gaps, including the nodes along the $k_x=\pm k_y$ lines where the orbital hybridization is forbidden, further supports this interpretation.

This mechanism differs fundamentally from approaches restricted to $\omega=0$ Fermi-surface properties, where the pairing symmetry is highly sensitive to the detailed Fermi-surface topology and nesting conditions.
In such frameworks, the $s_{\pm}$ state is typically stabilized by scattering among $\alpha$, $\beta$, and $\gamma$ Fermi-surface sheets and may rapidly evolve into a $d$-wave superconducting state once the $\gamma$ sheet disappears.
In contrast, the pairing state obtained here remains robust even when the $\gamma$ sheet vanishes, reflecting the fact that the pairing interaction originates from the bonding--antibonding $z^2$ channel rather than from the detailed topology of the Fermi surface.

Our analysis also highlights that superconducting correlations are governed not only by the pairing interaction but also by the orbital character of the low-energy electronic states.
Although the pairing interaction primarily resides in the $z^2$ channel, strong hybridization redistributes superconducting correlations to the $x^2-y^2$ channel, which contributes significantly to the low-energy quasiparticles.
Consistently, the orbital-resolved density of states near the Fermi level closely tracks the behavior of the superconducting correlations, explaining why $P_{xx}$ grows while $P_{zz}$ decreases but remains finite even after the $\gamma$ Fermi-surface sheet disappears.

Experimentally, superconductivity in La$_3$Ni$_2$O$_7$ emerges under high pressure and exhibits only a weak pressure dependence of $T_c$ once established.
Although a direct mapping between pressure and $\Delta E/t_\perp$ is beyond the scope of the present study, the robustness of superconductivity over a broad parameter range in our calculations—arising from orbital hybridiation and resulting redistribution—may provide a microscopic basis for this stability.

These results demonstrate the hierarchical structure of primary and hybridization-induced superconducting correlations, a distinctive feature of bilayer nickelates.
This perspective highlights the importance of orbital hierarchy and hybridization in multilayer correlated superconductors and provides guidance for exploring related nickelate and multiorbital superconducting systems.

\section*{Methods}

\subsection*{Bilayer two-orbital Hubbard model}
We consider a bilayer two-orbital Hubbard model with orbitals $m\in\{x,z\}$ (representing $x^2-y^2$ and $z^2$) and layer index $l\in\{1,2\}$.
The electron annihilation operator in the orbital/layer basis is denoted by $c_{i m_l\sigma}$, and its Fourier transform is
$c_{\mathbf{k}m_l\sigma} = N^{-1/2}\sum_i e^{-i\mathbf{k}\cdot\mathbf{R}_i} c_{i m_l\sigma}$,
where $N$ is the number of unit cells.

The kinetic energy term $H_{\text{kin}}$ is given by
\begin{align}
H_{\text{kin}}
&=\sum_{\mathbf{k},\sigma}
\left( c_{\mathbf{k}x_1\sigma}^{\dagger},\, c_{\mathbf{k}x_2\sigma}^{\dagger},\,  
c_{\mathbf{k}z_1\sigma}^{\dagger},\, c_{\mathbf{k}z_2\sigma}^{\dagger}\right)
\,\hat{T}(\mathbf{k})\,
\begin{pmatrix}
c_{\mathbf{k}x_1\sigma} \\ c_{\mathbf{k}x_2\sigma} \\ 
c_{\mathbf{k}z_1\sigma} \\ c_{\mathbf{k}z_2\sigma}
\end{pmatrix}.
\label{Hkin}
\end{align}
The matrix elements $T(\mathbf{k})$ include the hopping terms, the on-site level offset $\Delta E=\varepsilon_x-\varepsilon_z$, and the interorbital hybridization.
All hopping parameters are chosen to reproduce the band structure obtained from first-principles calculations.
The band structure for $\Delta E/t_{\perp}=0.50$ is shown in Fig.~\ref{band}.

Diagonalizing $\hat{T}(\mathbf{k})$ by a unitary transformation,
$\hat{U}^\dagger(\mathbf{k})\,\hat{T}(\mathbf{k})\,\hat{U}(\mathbf{k})=\mathrm{diag}\{E_\mu(\mathbf{k})\}$,
we introduce band operators $a_{\mathbf{k}\mu\sigma}$ via
$c_{\mathbf{k}\alpha\sigma}=\sum_{\mu}U_{\alpha\mu}(\mathbf{k})\,a_{\mathbf{k}\mu\sigma}$,
where $\alpha\in\{x_1,x_2,z_1,z_2\}$ labels the orbital/layer basis.
This yields
\begin{align}
H_{\text{kin}}=\sum_{\mathbf{k},\mu,\sigma}E_{\mu}(\mathbf{k})a^{\dagger}_{\mathbf{k}\mu\sigma}a_{\mathbf{k}\mu\sigma}.
\label{Hkin_diag}
\end{align}

The interaction term is given by the standard on-site multiorbital form,
\begin{align}
H_{\rm int}
&=U\sum_{i,m,l} n^{m_l}_{i\uparrow} n^{m_l}_{i\downarrow}
+\left(U'-\frac{J}{2}\right)\sum_{i,l} n^{x_l}_i\, n^{z_l}_i \notag\\
&-2J\sum_{i,l}\mathbf{S}^{x_l}_i\cdot\mathbf{S}^{z_l}_i -J'\sum_{i,l}\left(
c^{\dagger}_{ix_l\uparrow}c^{\dagger}_{ix_l\downarrow}c_{iz_l\downarrow}c_{iz_l\uparrow}
+\mathrm{H.c.}
\right).
\label{Hint}
\end{align}
Here $n^{m_l}_{i\sigma}=c^{\dagger}_{im_l\sigma}c_{im_l\sigma}$, $n^{m_l}_i=\sum_{\sigma}n^{m_l}_{i\sigma}$, and
$\mathbf{S}^{m_l}_i=\frac{1}{2}\sum_{\alpha\beta}c^{\dagger}_{im_l\alpha}\boldsymbol{\sigma}_{\alpha\beta}c_{im_l\beta}$.
We impose the rotationally invariant conditions $U'=U-2J$ and $J'=J$.

Finally, we introduce a double-counting correction to remove the static Hartree contributions already included in the reference band structure~\cite{Anisimov1991bta,Watanabe2021udo}. %~\cite{Anisimov1991bta,Hansmann2014iod,Watanabe2021udo} 
This is implemented as a shift of the orbital energies,
\begin{align}
\varepsilon_z &\rightarrow \varepsilon_z 
-\left[\frac{U}{2}n^z_0 
+\left(U'-\frac{J}{2}\right)n^x_0 \right], \notag \\
\varepsilon_x &\rightarrow \varepsilon_x 
-\left[\frac{U}{2}n^x_0 
+\left(U'-\frac{J}{2}\right)n^z_0 \right],
\end{align}
where $n^m_0$ denotes the orbital occupancy obtained from the noninteracting band structure.
This correction ensures that only correlation effects beyond the static mean-field contribution of the reference band structure are treated explicitly.
Note that for each value of $\Delta E$, the reference orbital occupancies $n_0^m(\Delta E)$ are recalculated from the corresponding noninteracting band structure and used to define the double-counting correction.
This ensures that the static Hartree contribution associated with the chosen bare orbital hierarchy is consistently removed.

The total Hamiltonian is a sum of the kinetic energy term with the double counting corrections and the interaction term,
\begin{equation}
    H=H^{\text{DCC}}_{\text{kin}}+H_{\text{int}}.
\end{equation}

\subsection*{Variational wave function}
The ground state is obtained within the variational Monte Carlo (VMC) framework.
We employ a Gutzwiller--Jastrow type trial wave function of the form
\begin{equation}
|\Psi\rangle = P^{(2)}_G P_{J_{\text{c}}} P_{J_{\text{s}}} |\Phi\rangle.
\end{equation}
$P^{(2)}_G$ is the Gutzwiller factor extended to the two-orbital system~\cite{Bunemann1998mgw,Watanabe2021udo}, assigning independent variational weights to the 16 possible local charge and spin configurations of the $z^2$ and $x^2-y^2$ orbitals at each site.
$P_{J_{\text{c}}}$ and $P_{J_{\text{s}}}$ are charge and spin Jastrow factors, respectively.
The charge Jastrow factor incorporates orbital-dependent density--density correlations, including both intra- and interorbital terms, while the spin Jastrow factor introduces correlations between the $z$ components of spin operators with orbital-dependent couplings.
$|\Phi\rangle$ is constructed as the ground state of an effective Bogoliubov--de Gennes (BdG) Hamiltonian.%~\cite{Himeda2002ssw}.

The BdG Hamiltonian is written as
\begin{equation}
H_{\text{BdG}}
=\sum_{i,j}
\Psi_i^\dagger
\begin{pmatrix}
\hat{T}_{ij}-\mu \hat{I} & \hat{\Delta}_{ij} \\
\hat{\Delta}_{ij}^\dagger & -\hat{T}_{ij}+\mu \hat{I}
\end{pmatrix}
\Psi_j,
\end{equation}
where $\Psi_i^\dagger=(c_{ix_1\uparrow}^\dagger,c_{ix_2\uparrow}^\dagger,
c_{iz_1\uparrow}^\dagger,c_{iz_2\uparrow}^\dagger,
c_{ix_1\downarrow},c_{ix_2\downarrow},
c_{iz_1\downarrow},c_{iz_2\downarrow})$
is the Nambu spinor.
Here $\hat{T}_{ij}$ is a $4\times4$ matrix in the orbital/layer space corresponding to the renormalized kinetic term and $\hat{\Delta}_{ij}$ represents the variational pairing fields.
The variational parameters in $|\Phi\rangle$ consist of renormalized hopping amplitudes, the pairing fields, and the chemical potential $\mu$, with $t_\perp$ fixed as the unit of energy.
Setting $\hat{\Delta}_{ij}=0$ yields the normal (paramagnetic) state.
This formulation allows, in principle, a smooth description from the BCS to the BEC regime. %~\cite{Watanabe2025pob}.
The resulting BdG ground state is projected onto the fixed particle-number sector with $N/2$ pairs corresponding to the target filling.

All variational parameters are optimized by minimizing the total energy using the stochastic reconfiguration method~\cite{Sorella2001gla}.
In particular, we focus on the interlayer pairing components within the same orbital, denoted as $\tilde{\Delta}_{zz}$ and $\tilde{\Delta}_{xx}$, which correspond to the optimized values of the variational pairing fields shown in Fig.~\ref{Delta}.
In addition to these dominant components, other symmetry-allowed pairing fields are included in the variational Hamiltonian.
Among them, a small intralayer component $\tilde{\Delta}^{\text{intra}}_{zz}$ appears, whose magnitude is typically about $10\%$ of the interlayer $\tilde{\Delta}_{zz}$.
We also include an intralayer-interorbital pairing component between the $z^2$ and $x^2-y^2$ orbitals with $d$-wave–like symmetry.
Although its optimized amplitude is nearly zero, this term is retained to preserve the symmetry of the BdG Hamiltonian and to avoid artificial decoupling between the orbital channels.

\subsection*{Superconducting correlation function}
The interlayer pair-correlation function within orbital $m$ is defined as
\begin{equation}
P_{mm}(\mathbf{r})
=
\frac{1}{N}
\sum_i
\langle
F^\dagger_{m_1m_2}(\mathbf{R}_i)
F_{m_1m_2}(\mathbf{R}_i+\mathbf{r})
\rangle,
\end{equation}
where
\begin{equation}
F^{\dagger}_{m_1m_2}(\mathbf{R}_i)
=
\frac{1}{\sqrt{2}}
\left(
c^{\dagger}_{im_1\uparrow} c^{\dagger}_{im_2\downarrow}
-
c^{\dagger}_{im_1\downarrow} c^{\dagger}_{im_2\uparrow}
\right)
\end{equation}
is an interlayer spin-singlet pair operator and $N$ is the number of unit cells.
We define the long-distance limit
\begin{equation}
P_{mm}=\lim_{|\mathbf{r}|\to\infty} P_{mm}(\mathbf{r}),
\end{equation}
which characterizes the strength of long-range superconducting correlations in orbital channel $m$.
This quantity serves as an indicator of superconducting order in each orbital channel.
In practice, the long-distance value $P_{mm}$ is evaluated by averaging $P_{mm}(\mathbf{r})$ over distances $6 \le |\mathbf{r}| \le L/2$ on a finite lattice of linear size $L=20$.
This procedure suppresses short-distance contributions and provides a stable estimate of the long-range superconducting correlations.

We have also examined intra- and interlayer pair-correlation functions corresponding to $d$-wave pairing symmetries proposed in previous studies and find that they do not develop sizable long-range correlations in the present parameter regime.

\subsection*{Orbital-resolved density of states}
To analyze the low-energy electronic structure, we calculate the orbital- and band-resolved density of states near the Fermi level.
For orbital $m$ and band $\mu$, it is defined as
\begin{equation}
D^{\mu}_{m}
=
\frac{1}{N_{\mathbf{k}}}
\sum_{\mathbf{k}}
|U_{m\mu}(\mathbf{k})|^2
\,\delta_\eta \!\left(E_\mu(\mathbf{k})-E_F\right),
\end{equation}
where $E_\mu(\mathbf{k})$ is the band dispersion, 
$U_{m\mu}(\mathbf{k})$ is the orbital weight of band $\mu$ obtained from diagonalizing the kinetic energy term of the Hamiltonian, 
and $N_{\mathbf{k}}$ is the number of $\mathbf{k}$ points.
The broadened delta function is taken as
\begin{equation}
\delta_\eta(x)=\frac{1}{\sqrt{\pi}\eta}e^{-(x/\eta)^2},
\end{equation}
with a finite energy width $\eta=0.05\,t_\perp$.
The finite broadening ensures that the density of states remains well defined even when the $\gamma$ Fermi-surface sheet vanishes.

The orbital-resolved density of states is obtained by summing over bands,
\begin{equation}
D_m=\sum_\mu D_m^\mu.
\end{equation}
In the main text, $D_{x^2-y^2}$ is dominated by contributions from the $\alpha$ and $\beta$ bands. 
In contrast, due to strong orbital hybridization, the $z^2$ character is not confined to the $\gamma$ band but is also substantially distributed over the $\alpha$ and $\beta$ bands.

\section{Data availability}
The data that support the findings of this study are available from the corresponding author upon reasonable request.

\section{Code availability}
The codes used for the VMC calculation are available from the corresponding author upon reasonable request.

\section{Acknowledgements}
We thank Dr. Y. Yamakawa for fruitful discussions.
This work was supported by JSPS KAKENHI Grant Nos. JP24K01333, JP25K08459, and JP25H01252, and by the Fusion Oriented Research for disruptive Science and Technology (FOREST) Program from the Japan Science and Technology Agency (JST), Grant No. JPMJFR246T.
The computing resource is supported by the supercomputer system (system-B) in the Institute for Solid State Physics, the University of Tokyo.

\section{Author contributions}
H.W. performed the variational Monte Carlo calculations.
H.S. performed the first-principles calculations and model construction.
K.K. supervised the project.
All authors discussed the results and contributed to writing the manuscript. 
\section{Competing interests}
The authors declare no competing interests.


\section{References}
\begin{thebibliography}{99} %% The number "99" means that this list has more than nine items.
\bibitem{Sun2023sos} Sun, H. et al. Signatures of superconductivity near 80 K in a nickelate under high pressure. \textit{Nature} \textbf{621}, 493 (2023).
\bibitem{Zhang2024hsw} Zhang, Y. et al. High-temperature superconductivity with zero resistance and strange-metal behaviour in La$_3$Ni$_2$O$_{7-\delta}$. \textit{Nat. Phys.} \textbf{20}, 1269 (2024).
\bibitem{Wang2024psi} Wang, G. et al. Pressure-Induced Superconductivity In Polycrystalline La$_3$Ni$_2$O$_{7-\delta}$. \textit{Phys. Rev. X} \textbf{14}, 011040 (2024).
\bibitem{Wang2024bhs} Wang, G. et al. Bulk high-temperature superconductivity in pressurized tetragonal La$_2$PrNi$_2$O$_7$. \textit{Nature} \textbf{634}, 579 (2024).
\bibitem{Li2026bsu} Li, F. et al. Bulk superconductivity up to 96 K in pressurized nickelate single crystals. \textit{Nature} \textbf{649}, 871 (2026).
\bibitem{Ko2025soa} Ko, E. K. et al. Signatures of ambient pressure superconductivity in thin film La$_3$Ni$_2$O$_7$. \textit{Nature} \textbf{638}, 935 (2025).
\bibitem{Zhou2025aso} Zhou, G. et al. Ambient-pressure superconductivity onset above 40K in (La,Pr)$_3$Ni$_2$O$_7$ films. \textit{Nature} \textbf{640}, 641 (2025). 
\begin{comment}
\bibitem{Zhou2025soa} Zhou, G. et al. Superconductivity onset above 60 K in ambient-pressure nickelate films. arXiv:2512.04708 .
\bibitem{Sakakibara2024tao} Sakakibara, H. et al. Theoretical analysis on the possibility of superconductivity in the trilayer Ruddlesden-Popper
nickelate La$_4$Ni$_3$O$_{10}$ under pressure and its experimental examination: Comparison with La$_3$Ni$_2$O$_7$. \textit{Phys. Rev. B} \textbf{109}, 144511 (2024). 
\bibitem{Li2024sos} Li, Q. et al. Signature of Superconductivity in Pressurized La$_4$Ni$_3$O$_{10}$. \textit{Chin. Phys. Lett.} \textbf{41}, 017401 (2024).
\bibitem{Zhu2024sip} Zhu, Y. et al. Superconductivity in pressurized trilayer La$_4$Ni$_3$O$_{10-\delta}$ single crystals. \textit{Nature} \textbf{631}, 531 (2024).
\bibitem{Zhang2025sit} Zhang, M. et al. Superconductivity in Trilayer Nickelate La$_4$Ni$_3$O$_{10}$ under Pressure. \textit{Phys. Rev. X} \textbf{15}, 021005 (2025).
\bibitem{Zhang2025bsi} Zhang, E. et al. Bulk Superconductivity in Pressurized Trilayer Nickelate Pr$_4$Ni$_3$O$_{10}$ Single Crystals. \textit{Phys. Rev. X} \textbf{15}, 021008 (2025).
\end{comment}
\bibitem{Shilenko2023ces} Shilenko, D. A. \& Leonov, I. V. Correlated electronic structure, orbital-selective behavior, and magnetic correlations in double-layer La$_3$Ni$_2$O$_7$ under pressure. \textit{Phys. Rev. B} \textbf{108}, 125105 (2023).
\bibitem{Luo2023btm} Luo, Z., Hu, X., Wang, M., W\'{u}, W. \& Yao, D.-X. Bilayer Two-Orbital Model of La$_3$Ni$_2$O$_7$ under Pressure. \textit{Phys. Rev. Lett.} \textbf{131}, 126001 (2023).
\bibitem{Christiansson2023ces} Christiansson, V., Petocchi, F. \& Werner, P. Correlated electronic structure of La$_3$Ni$_2$O$_7$ under pressure. \textit{Phys. Rev. Lett.} \textbf{131}, 206501
(2023).
\bibitem{Zhang2023esd} Zhang, Y., Lin, L.-F., Moreo, A. \& Dagotto, E. Electronic structure, dimer physics, orbital-selective behavior, and magnetic tendencies in the bilayer nickelate superconductor La$_3$Ni$_2$O$_7$ under pressure. \textit{Phys. Rev. B} \textbf{108}, L180510 (2023).
\bibitem{Cao2024fbp} Cao, Y. \& Yang, Y.-f. Flat bands promoted by Hund’s rule coupling in the candidate double-layer high-temperature superconductor La$_3$Ni$_2$O$_7$ under high pressure. \textit{Phys. Rev. B} \textbf{109}, L081105 (2024).
\bibitem{Wu2024sac} W\'{u}, W., Luo, Z., Yao, D.-X. \& Wang, M. Superexchange and charge transfer in the nickelate superconductor La$_3$Ni$_2$O$_7$ under pressure. \textit{Sci. China Phys. Mech. Astron.} \textbf{67}, 117402 (2024).
\bibitem{Rhodes2024srt} Rhodes, L. C. \& Wahl, P. Structural routes to stabilize superconducting La$_3$Ni$_2$O$_7$ at ambient pressure. \textit{Phys. Rev. Mater.} \textbf{8}, 044801 (2024).
\bibitem{Geisler2024sto} Geisler, B., Hamlin, J. J., Stewart, G. R., Hennig, R. G. \& Hirschfeld, P.
Structural transitions, octahedral rotations, and electronic properties of A$_3$Ni$_2$O$_7$ rare-earth nickelates under high pressure. \textit{npj Quantum Mater.} \textbf{9}, 38 (2024).
\bibitem{Wang2024eam} Wang, Y., Jiang, K., Wang, Z., Zhang, F.-C. \& Hu, J. Electronic and magnetic structures of bilayer La$_3$Ni$_2$O$_7$ at ambient pressure. \textit{Phys. Rev. B} \textbf{110}, 205122 (2024).
\bibitem{LaBollita2024asw} LaBollita, H., Pardo, V., Norman, M. R. \& Botana, A. S. Assessing spin-density wave formation in La$_3$Ni$_2$O$_7$ from electronic structure calculations. \textit{Phys. Rev. Mater.} \textbf{8}, L111801 (2024).
\bibitem{Nakata2017fsf} Nakata, M., Ogura, D., Usui, H. \& Kuroki, K. Finite-energy spin fluctuations as a pairing glue in systems with coexisting electron and hole bands. \textit{Phys. Rev. B} \textbf{95}, 214509 (2017).
\bibitem{Qin2023hsb} Qin, Q. \& Yang, Y.-f. High-$T_c$ superconductivity by mobilizing local spin singlets and possible route to higher $T_c$ in pressurized La$_3$Ni$_2$O$_7$. \textit{Phys. Rev. B} \textbf{108}, L140504 (2023).
\bibitem{Yang2023pss} Yang, Q.-G., Wang, D. \& Wang, Q.-H. Possible $s_{\pm}$-wave superconductivity in La$_3$Ni$_2$O$_7$. \textit{Phys. Rev. B} \textbf{108}, L140505 (2023).
\bibitem{Huang2023iav} Huang, J., Wang, Z. D. \& Zhou, T. Impurity and vortex states in the bilayer high-temperature superconductor La$_3$Ni$_2$O$_7$. \textit{Phys. Rev. B} \textbf{108}, 174501 (2023).
\bibitem{Shen2023ebm} Shen, Y., Qin, M. \& Zhang, G.-M. Effective Bi-Layer Model Hamiltonian and Density-Matrix Renormalization Group Study for the High-$T_c$ Superconductivity in La$_3$Ni$_2$O$_7$ under High Pressure. \textit{Chin. Phys. Lett.} \textbf{40}, 127401 (2023).
\bibitem{Oh2023ttm} Oh, H. \& Zhang, Y.-H. Type-II $t-J$ model and shared superexchange coupling from Hund’s rule in superconducting La$_3$Ni$_2$O$_7$. \textit{Phys. Rev. B} \textbf{108}, 174511 (2023).
\bibitem{Lechermann2023eca} Lechermann, F., Gondolf, J., B\"{o}tzel, S. \& Eremin, I. M. Electronic correlations and superconducting instability in La$_3$Ni$_2$O$_7$ under high pressure. \textit{Phys. Rev. B} \textbf{108}, L201121 (2023).
\bibitem{Liao2023eca} Liao, Z. et al. Electron correlations and superconductivity in La$_3$Ni$_2$O$_7$ under pressure tuning. \textit{Phys. Rev. B} \textbf{108}, 214522 (2023).
\bibitem{Qu2024btm} Qu, X.-Z. et al. Bilayer $t-J-J_{\perp}$ Model and Magnetically Mediated Pairing in the Pressurized Nickelate La$_3$Ni$_2$O$_7$. \textit{Phys. Rev. Lett.} \textbf{132}, 036502 (2024).
\bibitem{Kaneko2024pci} Kaneko, T., Sakakibara, H., Ochi, M. \& Kuroki, K. Pair correlations in the two-orbital hubbard ladder: Implications for superconductivity in the bilayer nickelate La$_3$Ni$_2$O$_7$. \textit{Phys. Rev. B} \textbf{109}, 045154 (2024).
\bibitem{Sakakibara2024pht} Sakakibara, H., Kitamine, N., Ochi, M. \& Kuroki, K. Possible High $T_c$ Superconductivity in La$_3$Ni$_2$O$_7$ under High Pressure through Manifestation of a Nearly Half-Filled Bilayer Hubbard Model. \textit{Phys. Rev. Lett.} \textbf{132}, 106002 (2024).
\bibitem{Heier2024cda} Heier, G., Park, K. \& Savrasov, S. Y. Competing $d_{xy}$ and $s_{\pm}$ pairing
symmetries in superconducting La$_3$Ni$_2$O$_7$ : LDA + FLEX calculations. \textit{Phys. Rev. B} \textbf{109}, 104508 (2024).
\bibitem{Zhang2024spt} Zhang, Y., Lin, L.-F., Moreo, A., Maier, T. A. \& Dagotto, E. Structural phase transition, $s_{\pm}$-wave pairing, and magnetic stripe order in bilayered superconductor La$_3$Ni$_2$O$_7$ under pressure. \textit{Nat. Commun.} \textbf{15}, 2470 (2024).
\bibitem{Lu2024ihs} Lu, C., Pan, Z., Yang, F., \& Wu, C. Interlayer-Coupling-Driven High-Temperature Superconductivity in La$_3$Ni$_2$O$_7$ under Pressure. \textit{Phys. Rev. Lett.} \textbf{132}, 146002 (2024).
\bibitem{Tian2024cea} Tian, Y. H. et al. Correlation effects and concomitant two-orbital $s_{\pm}$-wave superconductivity in La$_3$Ni$_2$O$_7$ under high pressure. \textit{Phys. Rev. B} \textbf{109}, 165154 (2024).
\bibitem{Botzel2024tom} B\"{o}tzel, S., Lechermann, F., Gondolf, J. \& Eremin, I. M. Theory of magnetic excitations in the multilayer nickelate superconductor La$_3$Ni$_2$O$_7$. \textit{Phys. Rev. B} \textbf{109}, L180502 (2024).
\bibitem{Kakoi2024pco} Kakoi, M., Kaneko, T., Sakakibara, H., Ochi, M. \& Kuroki, K. Pair correlations of the hybridized orbitals in a ladder model for the bilayer nickelate La$_3$Ni$_2$O$_7$. \textit{Phys. Rev. B} \textbf{109}, L201124 (2024).
\bibitem{Pan2024eor} Pan, Z., Lu, C., Yang, F. \& Wu, C. Effect of Rare-Earth Element Substitution in Superconducting R$_3$Ni$_2$O$_7$ under pressure. \textit{Chin. Phys. Lett.} \textbf{41}, 087401 (2024).
\bibitem{Chen2024osi} Chen, J., Yang, F. \& Li, W. Orbital-selective superconductivity in the pressurized bilayer nickelate La$_3$Ni$_2$O$_7$ : An inﬁnite projected entangled-pair state study. \textit{Phys. Rev. B} \textbf{110}, L041111 (2024).
\bibitem{Luo2024hsi} Luo, Z., Lv, B., Wang, M., W\'{u}, W. \& Yao, D.-X. High-$T_c$ superconductivity in La$_3$Ni$_2$O$_7$ based on the bilayer two-orbital $t-J$ model. \textit{npj Quantum Mater.} \textbf{9}, 61 (2024).
\bibitem{Ryee2024qpb} Ryee, S., Witt, N. \& Wehling, T. O. Quenched Pair Breaking by Interlayer Correlations as a Key to Superconductivity in La$_3$Ni$_2$O$_7$. \textit{Phys. Rev. Lett.} \textbf{133}, 096002 (2024).
\bibitem{Zhang2024spo} Zhang, J.-X., Zhang, H.-K., You, Y.-Z. \& Weng, Z.-Y. Strong pairing originated from an emergent ${\mathbb{Z}}_{2}$ berry phase in La$_3$Ni$_2$O$_7$. \textit{Phys. Rev. Lett.} \textbf{133}, 126501 (2024).
\bibitem{Yang2024spf} Yang, H., Oh, H. \& Zhang, Y.-H. Strong pairing from a small Fermi surface beyond weak coupling: Application to La$_3$Ni$_2$O$_7$. \textit{Phys. Rev. B} \textbf{110}, 104517 (2024).
\bibitem{Schlomer2024sit} Schl\"{o}mer, H., Schollw\"{o}ck, U., Grusdt, F. \& Bohrdt, A. Superconductivity in the pressurized nickelate La$_3$Ni$_2$O$_7$ in the vicinity of a BEC-BCS crossover. \textit{Commun. Phys.} \textbf{7}, 366 (2024).
\bibitem{Xia2025sdo} Xia, C., Liu, H., Zhou, S. \& Chen, H. Sensitive dependence of pairing symmetry on Ni-$e_g$ crystal ﬁeld splitting in the nickelate superconductor La$_3$Ni$_2$O$_7$. \textit{Nat. Commun.} \textbf{16}, 1054 (2025).
\bibitem{Jiang2025top} Jiang, K.-Y., Cao, Y.-H., Yang, Q.-G., Lu, H.-Y. \& Wang, Q.-H. Theory of pressure dependence of superconductivity in bilayer
nickelate La$_3$Ni$_2$O$_7$. \textit{Phys. Rev. Lett.} \textbf{134}, 076001 (2025).
\bibitem{Gu2025ema}  Gu, Y., Le, C., Yang, Z., Wu, X. \& Hu, J. Effective model and pairing
tendency in the bilayer Ni-based superconductor La$_3$Ni$_2$O$_7$. \textit{Phys. Rev. B} \textbf{111}, 174506 (2025).
\bibitem{Liu2025vqm} Liu, Y.-Q., Wang, D. \& Wang, Q.-H. Variational quantum Monte Carlo investigations of the superconducting pairing in La$_3$Ni$_2$O$_7$. \textit{Phys. Rev. B} \textbf{112}, 014511 (2025).
\bibitem{Qu2025hri} Qu, X.-Z., Qu, D.-W., Yi, X.-W., Li, W. \& Su, G. Hund’s rule, interorbital hybridization, and high-$T_c$ superconductivity
in the bilayer nickelate La$_3$Ni$_2$O$_7$. \textit{Phys. Rev. B} \textbf{112}, L161101 (2025).
\bibitem{Ryee2025sgb} Ryee, S., Witt, N., Sangiovanni, G. \& Wehling, T. O. Superconductivity Governed by Janus-Faced Fermiology in Strained Bilayer Nickelates. \textit{Phys. Rev. Lett.} \textbf{135}, 236003 (2025).
\bibitem{Gao2026rsp} Gao, Y. Robust $s_{\pm}$-wave pairing in a bilayer two-orbital model of pressurized La$_3$Ni$_2$O$_7$ without the $\gamma$ Fermi surface. \textit{Physica C} \textbf{640}, 1354824 (2026).
\bibitem{Inoue2026umo} Inoue, D., Yamakawa, Y., Onari, S. \& Kontani, H. Unified mechanism of charge-density-wave and high-$T_c$ superconductivity protected from oxygen vacancies in bilayer nickelates. \textit{Commun. Phys.} (2026).
\bibitem{Le2025osa} Le, C., Zhan, J., Wu, X. \& Hu, J. Opposite-Mirror-Parity Scattering as the Origin of Superconductivity in Strained Bilayer Nickelates. https://arxiv.org/abs/2501.14665~(2025).
\bibitem{Ushio2025tso} Ushio, K. et al. Theoretical study on ambient pressure superconductivity in La$_3$Ni$_2$O$_7$ thin films: structural analysis, model construction, and robustness of $s_{\pm}$-wave pairing. https://arxiv.org/abs/2506.20497 (2025).
\bibitem{Maier2025ipi} Maier, T. A. et al. Interlayer Pairing in Bilayer Nickelates. https://arxiv.org/abs/2506.07741 (2025).
\bibitem{Kuroki2002hsi} Kuroki, K., Kimura, T. \& Arita, R. High-temperature superconductivity in dimer array systems. \textit{Phys. Rev. B} \textbf{66}, 184508 (2002). 
\bibitem{Maier2011psa} Maier, T. A. \& Scalapino, D. J. Pair structure and the pairing interaction in a bilayer Hubbard model for unconventional superconductivity. \textit{Phys. Rev. B} \textbf{84}, 180513(R) (2011).
\bibitem{Nomura2025shs} Nomura, Y., Kitatani, M., Sakai, S. \& Arita, R. Strong-coupling high-$T_c$ superconductivity in doped correlated band insulators. \textit{Phys. Rev. B} \textbf{112}, L020504 (2025).
\bibitem{Jiang2024hsi} Jiang, K., Wang, Z. \& Zhang, F.-C. High-temperature superconductivity in La$_3$Ni$_2$O$_7$. \textit{Chin. Phys. Lett.} \textbf{41}, 017402 (2024).
\bibitem{Fan2024sin} Fan, Z. et al. Superconductivity in nickelate and cuprate superconductors with strong bilayer coupling. \textit{Phys. Rev. B} \textbf{110}, 024514 (2024).
\bibitem{Dagotto1992sil} Dagotto, E., Riera, J. \& Scalapino, D. Superconductivity in ladders and coupled planes. \textit{Phys. Rev. B} \textbf{45}, 5744 (1992).
\bibitem{Wang2025fla} Wang, J. \& Yang, Y.-F. Fermi liquid and isotropic superconductivity of Hund scenario for bilayer nickelates. https://arxiv.org/abs/2507.19301 (2025).
\bibitem{Yang2024oec} Yang, J. et al. Orbital-dependent electron correlation in double-layer nickelate La$_3$Ni$_2$O$_7$. \textit{Nat. Commun.} \textbf{15}, 4373 (2024).
\bibitem{Li2025aps} Li, P. et al. Angle-resolved photoemission spectroscopy of superconducting (La,Pr)$_3$Ni$_2$O$_7$/SrLaAlO$_4$ heterostructures. \textit{Natl. Sci. Rev.} \textbf{12}, nwaf205 (2025).
\bibitem{Wang2025eso} Wang, B. Y. et al. Electronic structure of compressively strained thin film La$_2$PrNi$_2$O$_7$. https://arxiv.org/abs/2504.16372 (2025).
\bibitem{Sun2025oos} Sun, W. et al. Observation of superconductivity-induced leading-edge gap in Sr-doped La$_3$Ni$_2$O$_7$ thin films. https://arxiv.org/abs/2507.07409 (2025).
\bibitem{Ochi2025tso} Ochi, M., Sakakibara, H., Usui, H. \& Kuroki, K. Theoretical study of the crystal structure of the bilayer nickel oxychloride Sr$_3$Ni$_2$O$_5$Cl$_2$ and analysis of possible unconventional superconductivity. \textit{Phys. Rev. B} \textbf{111}, 064511 (2025).
\bibitem{Ceperley1977mcs} Ceperley, D., Chester, G. V. \& Kalos, M. H. Monte Carlo simulation of a many-fermion study. \textit{Phys. Rev. B} \textbf{16}, 3081 (1977).
\bibitem{Yokoyama1987vms} Yokoyama, H. \& Shiba, H. Variational Monte-Carlo studies of Hubbard model. I. \textit{J. Phys. Soc. Japan} \textbf{56}, 1490 (1987).
\bibitem{Anisimov1991bta} Anisimov, V. I., Zaanen, J. \& Andersen, O. K. Band theory and Mott insulators: Hubbard $U$ instead of Stoner $I$. \textit{Phys. Rev. B} \textbf{44}, 943 (1991).
%\bibitem{Hansmann2014iod} Hansmann, P., Parragh, N., Toschi, A., Sangiovanni, G. \& Held, K. Importance of $d-p$ Coulomb interaction for high T$_c$ cuprates and other oxides. \textit{New J. Phys.} \textbf{16}, 033009 (2014).
\bibitem{Watanabe2021udo} Watanabe, H., Shirakawa, T., Seki, K., Sakakibara, H., Kotani, T., Ikeda, H. \& Yunoki, S. Unified description of cuprate superconductors using a four-band $d-p$ model. \textit{Phys. Rev. Res.} \textbf{3}, 033157 (2021).
\bibitem{Bunemann1998mgw} B\"{u}nemann, J., Weber, W. \& Gebhard, F. Multiband Gutzwiller wave functions for general on-site interactions \textit{Phys. Rev. B} \textbf{57}, 6896 (1998).
%\bibitem{Himeda2002ssw} Himeda, A., Kato, T. \& Ogata, M. Stripe states with spatially oscillating $d$-wave superconductivity in the two-dimensional $t-t'-J$ model. \textit{Phys. Rev. Lett.} \textbf{88}, 117001 (2002).
%\bibitem{Watanabe2025pob} Watanabe, H. \& Ikeda, H. Possibility of BCS-BEC crossover in $\kappa$-type organic superconductors. \textit{Phys. Rev. B} \textbf{111}, 085130 (2025).
\bibitem{Sorella2001gla} Sorella, S. Generalized Lanczos algorithm for variational quantum Monte Carlo \textit{Phys. Rev. B} \textbf{64}, 024512 (2001).
\end{thebibliography}
\end{document}